\documentclass[printer]{aa}

\usepackage{caption}
\usepackage{graphicx}
\usepackage[position=top]{subfig}
\usepackage{amsmath}
\usepackage[utf8]{inputenc}
\usepackage{epsfig,amsmath,amssymb}
\usepackage[varg]{txfonts}
\usepackage{academicons}
\usepackage[breaklinks, colorlinks, citecolor=blue]{hyperref}

\begin{document}

\title{The dynamical origin of the magnetic field distributions in compressible turbulence}
\authorrunning{Ntormousi and Del Sordo}
\titlerunning{Turbulent magnetic field distributions}
\author{Evangelia Ntormousi \orcid{0000-0002-4324-0034}\inst{1} and Fabio Del Sordo\orcid{0000-0001-9268-4849} \inst{1}}
\date{Received -- / Accepted --}

\institute{
Scuola Normale Superiore,
Piazza dei Cavalieri, 7
56126 Pisa, Italy
}

\abstract
{Magnetohydrodynamical (MHD) simulations of isothermal compressible turbulence report that the density distribution
is well described by a lognormal with a variance proportional to the flow's Mach number. 
The distribution of magnetic field strength 
also has
a lognormal component, but includes long, power-law-like tails. 
In this work, we use semi-analytical arguments to predict the distributions of density and magnetic field strength in compressible turbulent flows.
Specifically, in the Lagrangian description of the continuity and the induction equations, we model the velocity gradients of the turbulent flow as a simple random process, essentially turning these equations into stochastic differential equations. Integrating them leads to a lognormal distribution for the density field and the strength of the magnetic field. 
The power-law tails in the magnetic field PDF appear when we introduce intermittent shocks due to sampling rare events. Gradually increasing the frequency of these events, essentially going closer to a continuous process, leads to lognormal-like distributions again.
The asymmetry is connected to the relative abundance of slow and fast shocks. An overabundance of fast MHD shocks produces a high-value tail, while the contrary produces low-value tails. We propose that the appearance of power-law tails along lognormals in turbulent flows is the signature of the co-existence of continuous, diffusion-like propagation combined with localized, intermittent events.}


\maketitle
\section{Introduction}
The density and magnetic field distributions in turbulent astrophysical flows are key to understanding processes like star formation, since molecular clouds are subject to supersonic magnetohydrodynamic (MHD) turbulence \citep[e.g,][]{MacLow_klessen2004,Elmegreen_scalo2004,girichidis2020}. 
In this situation the density distribution provides the initial conditions for star formation, while the magnetic field distribution can be an indication of the importance of magnetic support \citep{Hennebelle_Inutsuka2019,pattle2023}. The description of compressible turbulent MHD  systems is often reduced to the study of the density and the magnetic field probability density functions (PDFs) as a consequence of their complexity.

To predict the PDFs of the gas density in turbulent environments, numerous studies have modeled hydrodynamic and MHD turbulence under different conditions  \citep[e.g.,][for an early review]{MacLow_klessen2004}.  
A well-established result is that the density in supersonic turbulence follows a lognormal distribution \citep[e.g.,][]{Vazquez-Semadeni1994,padoan1997,passot_vs1998,kowal2007,federrath2008,Federrath2010}, with variance $\rm \sigma^2 \sim b^2 \mathcal{M}^2 $, where $ \mathcal{M} $ is the Mach number and b is a factor depending on the type of forcing. 
Deviations from a log-normal shape have been observed in higher moments of the distribution, often attributed to intermittent dissipative structures such as shocks \citep[e.g.,][]{Federrath2010,Rabatin_collins2023}.

The PDF of the magnetic field is typically found to have a lognormal contribution, accompanied by exponential tails (power-laws in log-space). \citet{Schekochihin_2004} and \citet{Seta_2020} reported these power-law segments in the high-value side of the distribution, including cases with viscosity or magnetic diffusivity. 
\citet{Beatie2020} studied the PDF of the magnetic field fluctuations parallel and perpendicular to the mean field and also found that they deviate from a Gaussian, forming strong anisotropic tails. In  \citet{Seta_2021}, these tails appear in the magnetic field PDFs in subsonic and supersonic turbulence, both in the kinematic and saturated dynamo regimes, although the asymmetry (high-versus low-value excess) varies from one regime to another. Finally, \citet{ntormousi2024} report a lognormal PDF with high-value tails for the magnetic field fluctuations in multi-phase galaxy simulations with gravity.
So far, no theoretical model exists for the origin of these low-value tails. 

We propose such a model in this work. We first show that the log-normality of the density PDF  
is a direct consequence of the turbulence driving method, which is in effect time-correlated noise 
\citep{Scannapieco_2024}. 
Then, we argue that the lognormal component of the magnetic field strength PDF in driven turbulence has a similar origin, while we attribute the appearance of asymmetric tails to occasional, discontinuous jumps that correspond to shocks.

\section{Stochastic model for the density evolution}
\label{sec:density_model}

We first consider the Lagrangian formulation of the continuity equation:

\begin{equation}
\rm
    \frac{d\rho}{dt} +\rho\nabla\cdot{\bf{v}}=0. \label{eq:continuity} 
\end{equation}
We notice that, if $\rm \nabla\cdot\bf{v}$ is modeled as a random process, Eq. \ref{eq:continuity} becomes a stochastic differential equation (SDE).

Then Eq. \ref{eq:continuity} takes the form of a multiplicative SDE:
\begin{equation}
 \rm   d\rho = c~\rho~dW(t),
    \label{eq:mult_noise}
\end{equation}
where c is a constant with units of velocity gradient. Eq. \ref{eq:mult_noise} can be exactly solved by changing coordinates to $\rm y=\log~\rho$ so that
\begin{equation}
\rm
    dy=\frac{1}{\rho}d\rho-\frac{1}{2\rho^2}d\rho^2.
    \label{eq:newcoords}
\end{equation}
Then Eq. \ref{eq:newcoords} is directly integrable using Ito's formula, yielding \citep{gardinerbook} :
\begin{equation}    \rm y(t)=y(t_0)+c\left[W(t)-W(t_0)\right]-\frac{1}{2}c^2(t-t_0).
\label{eq:gensolution}
\end{equation}
where $\rm t_0$ is a reference initial time.
Since the intervals $\rm \left[W(t)-W(t_0)\right]$ are independent random variables, we see that if we approximate the turbulent velocity gradient as a Wiener process, the density distribution should be log-normal. 

The simple argument above appears in \citet{ColesJones1991} in the context of cosmological density perturbations.  \citet{Vazquez-Semadeni1994,passot_vs1998} also predicted a lognormal PDF for the density in compressible turbulence, applying the central limit theorem to the random density jumps encountered by a fluid element. Here, we reach the same conclusion by modeling the divergence of v as a continuous random variable and integrating the resulting SDE.
The main difference with these previous works is the connection with the velocity field. We see that, if we model $\rm \nabla\cdot\mathbf{v}$ as a random field, which is a crude approximation to what numerical simulations of turbulence in a box adopt, then the density PDF will be lognormal.

\section{Stochastic model for the magnetic field evolution}

We can also cast the Lagrangian form of the induction equation:
\begin{equation}
\rm
\frac{d \mathbf{B}}{dt} = (\mathbf{B}\cdot\nabla) \, \mathbf{v} - \mathbf{B}(\nabla \cdot \mathbf{v}). \
\label{eq:induction}
\end{equation}
as a system of SDEs by defining the tensor $\rm \mathbf{A} \equiv \nabla \boldsymbol{v}$ and introducing the unit vector $\rm \hat{\boldsymbol{B}} = \boldsymbol{B}/|\boldsymbol{B}|$, so that
\begin{align}
\rm
\frac{d \hat{\mathbf{B}}}{dt} &= (\mathbf{I} - \hat{\mathbf{B}}\hat{\mathbf{B}}^{\top}) \, \mathbf{A} \, \hat{\mathbf{B}}, \label{eq:Bhat} \\
\rm
\frac{d}{dt} \ln |\boldsymbol{B}| &= \hat{\boldsymbol{B}}^{\top} \mathbf{A} \, \hat{\boldsymbol{B}} - \mathrm{tr}(\mathbf{A}). \label{eq:logB}
\end{align}
Equation~\eqref{eq:Bhat} describes the rotation and stretching of the field direction, while Eq.~\eqref{eq:logB} captures the amplification or decay of the field magnitude due to stretching along $\hat{\boldsymbol{B}}$ and compression or expansion of the flow.

For consistency, this model also propagates the density, following Eq. \ref{eq:continuity} with $\rm \nabla\cdot v\equiv tr(\textbf{A})$. 

\subsection{Stochastic modeling of the velocity gradient}

$\mathbf{A}$ is composed by a symmetric and an antisymmetric part:
\begin{equation}
\mathbf{A} = \mathbf{S} + \mathbf{\Omega},
\end{equation}
where $\mathbf{S}$ is the strain tensor and $\mathbf{\Omega}$ the rotation tensor.

We construct the strain tensor from three eigenvalues $\{\lambda_i\}$ and a random orthogonal eigenbasis drawn from the Haar measure. The eigenvalues are sampled as Gaussian random variables with variance $\sigma_\lambda^2$ and shifted such that
$\rm \sum_i \lambda_i = \theta, $
where $\rm \theta = \mathrm{tr}(\mathbf{A})$ is the divergence.
$\theta$ is drawn from a Gaussian distribution with variance
\begin{equation}
\rm 
\mathrm{Var}(\theta)
= C_{\rm bg}
\left(
\frac{v_{\rm rms}}{\ell_{\rm int}}
\right)^2.
\label{eq:vartheta}
\end{equation}
where $C_{\rm bg}$ is a dimensionless scaling constant that relates to the driving scale of the turbulence, here assumed equal to one.
The value of $\rm v_{rms}$ is directly calculated from the rms Mach number we assume for each model, $\rm v_{rms}=c_s~\mathcal{M}$, where the sound speed is set to unity for all models. In this approach there is the implicit assumption that the flow is statistically steady (e.g., driven turbulence) so that $\rm \mathbf{A}$ stays random with the same statistical behavior on the timescales in which we are interested. This is equivalent to driven turbulence, but not to time-dependent flows like, e.g., decaying turbulence.
The rotation tensor is constructed by subtracting a random matrix from its transpose and multiplying by one half.

So far, the model evolves the density purely through these continuous random fluctuations of the velocity divergence, so we expect the density PDF to be a log-normal, as predicted by the analytical argument in Sec. \ref{sec:density_model}. The magnetic field, however, could already be subject to more complex effects like stretching and compression through the first term in Eq. \ref{eq:logB}.

In addition to the stochastic strain dynamics, we have included a phenomenological alignment term that models the tendency of magnetic field lines to align with particular strain eigenvectors in turbulent flows.
The target eigenvector may correspond to the compressive, stretching, or intermediate eigenvalue direction, allowing exploration of different alignment regimes. However, changes in this alignment do not alter the resulting PDFs, so all the results will be presented for the intermediate eigenvalue direction.

\begin{figure}
    \centering
    \includegraphics[width=0.85\linewidth]{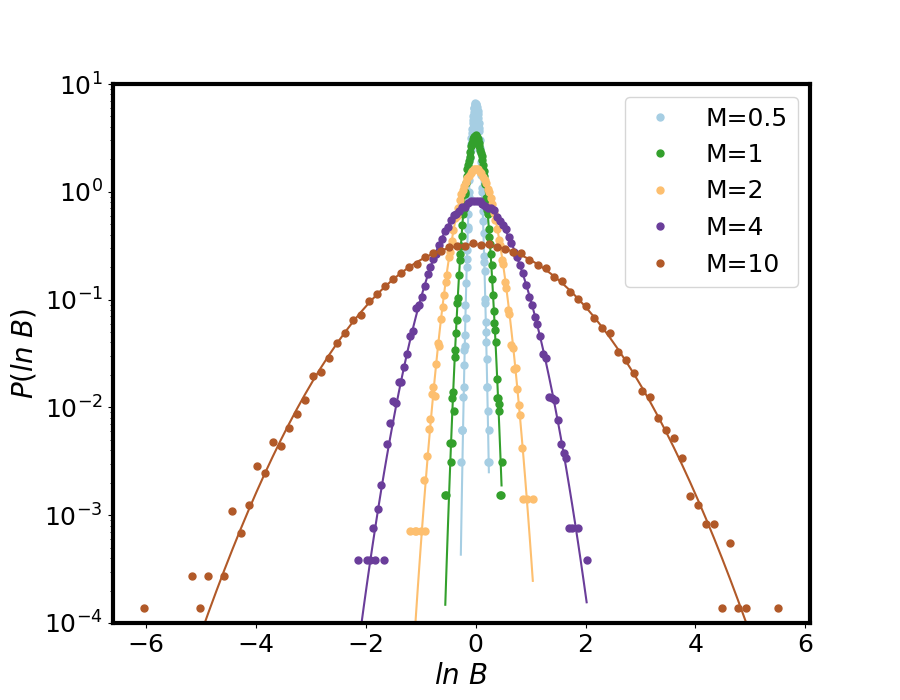}
     \includegraphics[width=0.85\linewidth]{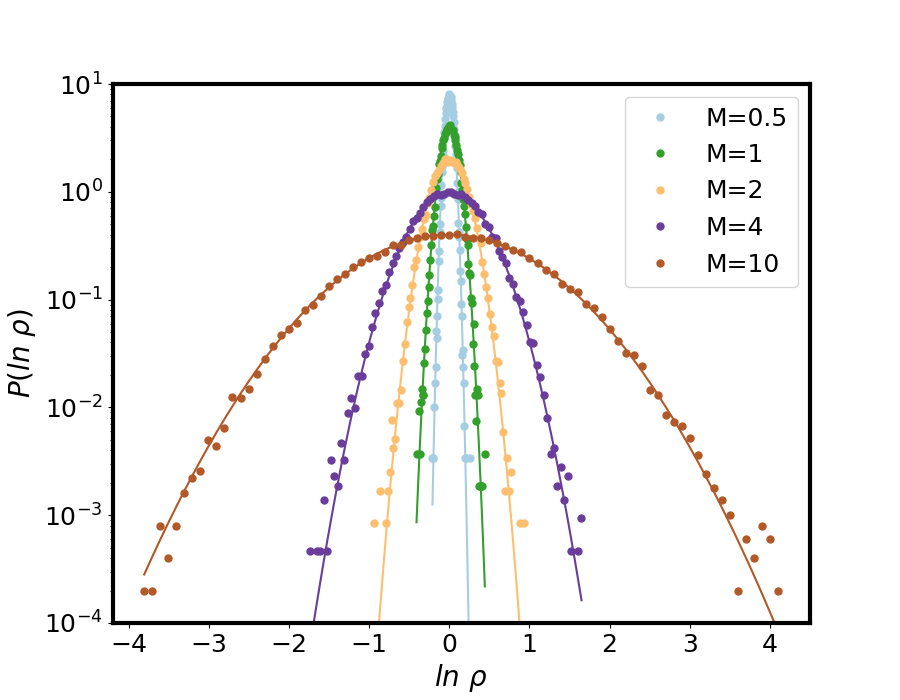}
    \caption{Magnetic field and density PDFs in the absence of shocks ($\rm p_{shock}=0$), for different Mach numbers.}
    \label{fig:noshocks}
\end{figure}

\subsection{Intermittent shock events}

To model intermittency, we introduce discrete jump events along each trajectory, representing shock passages. These occur as a Poisson process with probability $\rm p_{shock}$ per timestep.
For each shock event, we draw a dimensionless compression ratio
$\rm r > 1 $ from a heavy-tailed distribution,
\begin{equation}
\rm
r \sim r_0\, U^{-1/\alpha},
\label{eq:shock_mag}
\end{equation}
where $\rm U\in(0,1)$ is uniform and $\alpha$ controls the tail strength.
In principle, $\alpha$ can have any positive value. However, since
\citet{Smith2000} find $\alpha\simeq2$ in simulations of hydrodynamic turbulence, we consider this to be the fiducial value. In practice, we also truncate r to avoid unphysically large variance.

Across each shock, the density undergoes an instantaneous Rankine-Hugoniot (RH) jump,
$\rm \Delta \ln \rho_{comp} \equiv \ln r $.
To represent post-shock expansion, each compressive jump is followed by a rarefaction event of reduced magnitude,
$\rm \Delta \ln \rho_{\rm rare} = -\kappa \ln r$, with $\rm 0 < \kappa \le 1$. The parameter $\rm \kappa$ controls the relative strength of expansion compared to compression and is here set to $\rm \kappa=0.25$. 

The RH conditions are also applied to the magnetic field. If we decompose $\rm \mathbf{B}$ into the normal and tangential components with respect to the shock normal, $\rm \mathbf{B_n}$ and $\rm \mathbf{B}_t$, $\rm \mathbf{B_n}$ remains continuous across the shock, but $\rm \mathbf{B_t}$ is rescaled, $\rm \mathbf{B}_{t,2} = s\, \mathbf{B}_{t,1}$.
The scaling factor $\rm s$ distinguishes fast and slow shocks:
For a fast shock $\rm s = r $, while for a slow shock $\rm s = r^{-1}$.
In other words, fast shocks amplify the tangential magnetic field, whereas slow shocks reduce it.

The fraction of fast shocks, and therefore the probability that the encountered shock is fast (i.e. causes magnetic field amplification rather than decrease) is controlled by the parameter q.
We ignore intermediate shocks here, so the percentage of slow shocks will be $\rm 1-q$. In physical systems,
q is expected to depend on the Alfv\'{e}nic Mach number and plasma $\beta$, but here we treat it as free, because the relative abundance of slow and fast shocks in MHD turbulence is far from understood. Notably, \citet{Lehmann2016} found a significant overabundance of slow shocks in one case of high Mach turbulence, but their analysis was not complete over Mach and Alfv\'{e}nic Mach numbers. This finding implies that the parameter q in our model can be very close to zero in some situations. 

\section{Analytical insight into the appearance of the exponential tails}

Before proceeding to integrate the above model numerically, it is useful to look into the analytical reasons why we expect exponential tails to arise in this situation.
By adding the occasional jumps in the model, we are introducing a Poisson process to the continuous, diffusion-like process represented by the random velocity gradient.
We can write the jump-diffusion process as an SDE:
\begin{equation}
\rm    
dX=\mu\,dt+\sigma\,dW_t-D\,dN_t ,
\end{equation}
where $\rm W_t$ is the Wiener process, $\rm N_t$ is a Poisson process with rate $\lambda$, and D is the logarithmic decrement associated with a single shock.  If, as we have set in our model, the shock multiplier is a power-law function of the form
$ \rm 
r=r_0 U^{1/\alpha}$ 
then 
$ \rm D=-\ln r=-\ln r_0-\frac{1}{\alpha}\ln U$.
Since the the random variable produced by taking the logarithm of a uniform distribution will be exponentially distributed, one may write
$\rm D=c+Y$, where $\rm c=-\ln r_0$ and $ \rm Y\sim \mathrm{Exp}(\alpha)$
where \(\alpha\), our parameter for the slope of shock strength distribution, becomes the exponential rate.
The integrated process is therefore:
\begin{equation}
\rm    
X(t)=X_0+\mu t+\sigma W_t-\sum_{i=1}^{N_t}D_i,
\end{equation}
with $\rm \{D_i\}_{i\geq 1}$ independent copies of D, independent of $\rm W_t$ and $\rm N_t$.  Conditioning on the number of shocks, $\rm N_t=n$, gives:
\begin{equation}
\rm    
X(t)\mid N_t=n
=
X_0+\mu t+\sigma W_t-nc-\sum_{i=1}^{n}Y_i .
\end{equation}
Because $\rm Y_i$ are independent exponential random variables with common rate $\alpha$, their sum is gamma distributed:
\begin{equation}
\rm    
\sum_{i=1}^{n}Y_i \sim \Gamma(n,\alpha),
\end{equation}
where the first argument is the shape and the second is the rate.  Hence, for $\rm n\geq 1$,
\begin{equation}
\rm
X(t)\mid N_t=n
=
m_t+\sigma W_t-nc-G_n,
\end{equation}
where $\rm m_t=X_0+\mu t$ and $\rm G_n\sim \Gamma(n,\alpha)$.
The corresponding conditional density is the convolution of a Gaussian density with a shifted gamma law:
\begin{equation}
\rm    
p_n(x,t)
=
\int_0^\infty
\frac{1}{\sqrt{2\pi\sigma^2 t}}
\exp\!\left[
-\frac{(x-m_t+nc+g)^2}{2\sigma^2 t}
\right]
\frac{\alpha^n}{\Gamma(n)}g^{n-1}e^{-\alpha g}\,dg .
\end{equation}
For $\rm n=0$, the gamma contribution is absent and we recover the lognormal distribution for the magnetic field strength.
The gamma distribution enters the conditional PDF as the law of the accumulated logarithmic shock decrement at fixed shock count. It persists as long as the added shocks have an exponential or $\rm \chi^2$ distribution.

\begin{figure*}
    \centering
    \includegraphics[trim={0cm 0cm 1cm 1cm}, clip,width=0.33\linewidth]{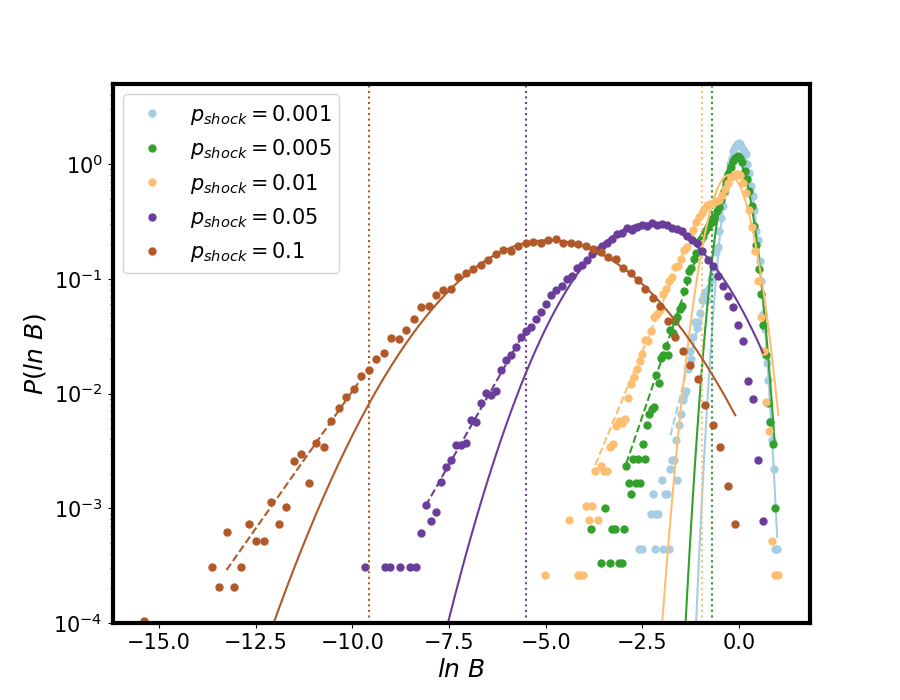}
    \includegraphics[trim={0cm 0cm 1cm 1cm},clip ,width=0.33\linewidth]{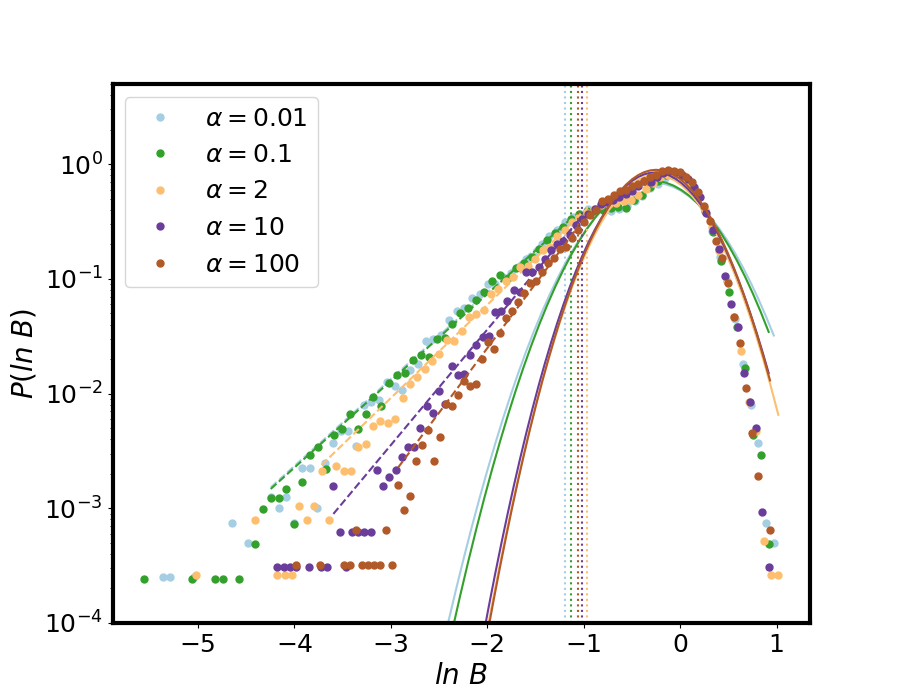}
    \includegraphics[trim={0cm 0cm 1cm 1cm},clip ,width=0.33\linewidth]{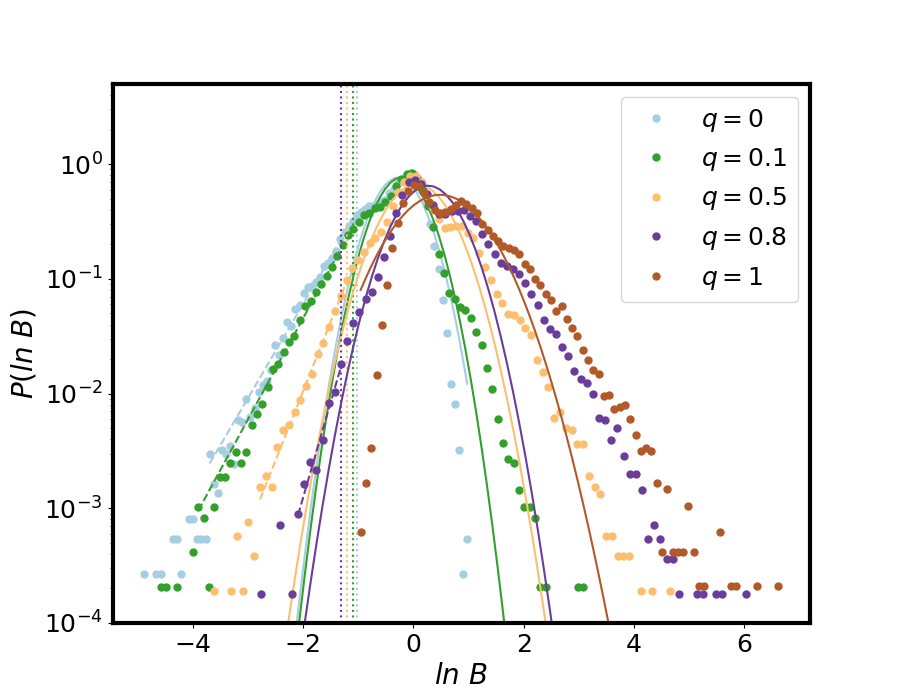}
    \includegraphics[trim={0cm 0cm 1cm 1cm}, clip,width=0.33\linewidth]{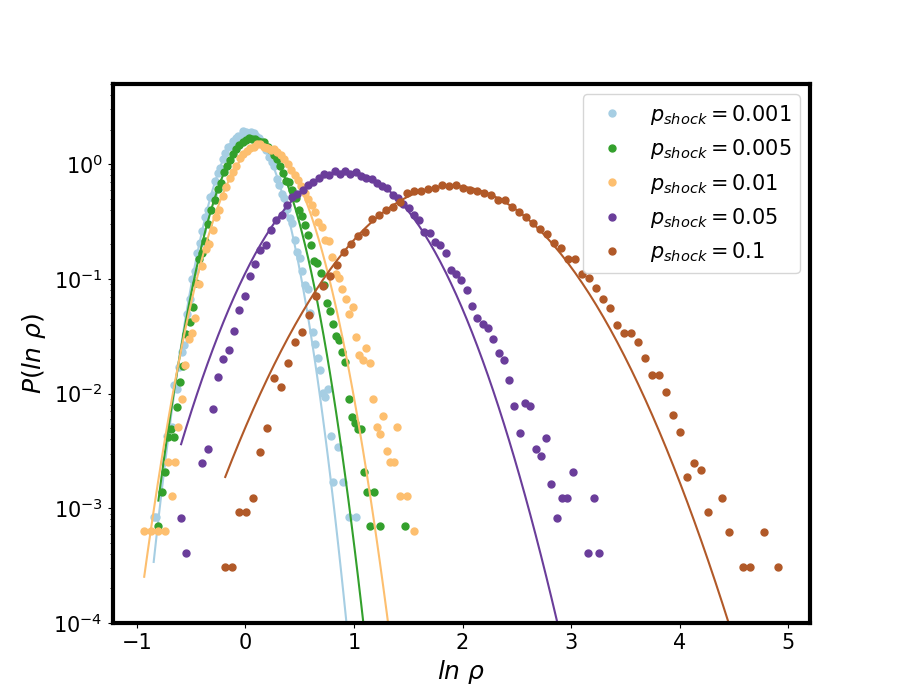}
    \includegraphics[trim={0cm 0cm 1cm 1cm},clip ,width=0.33\linewidth]{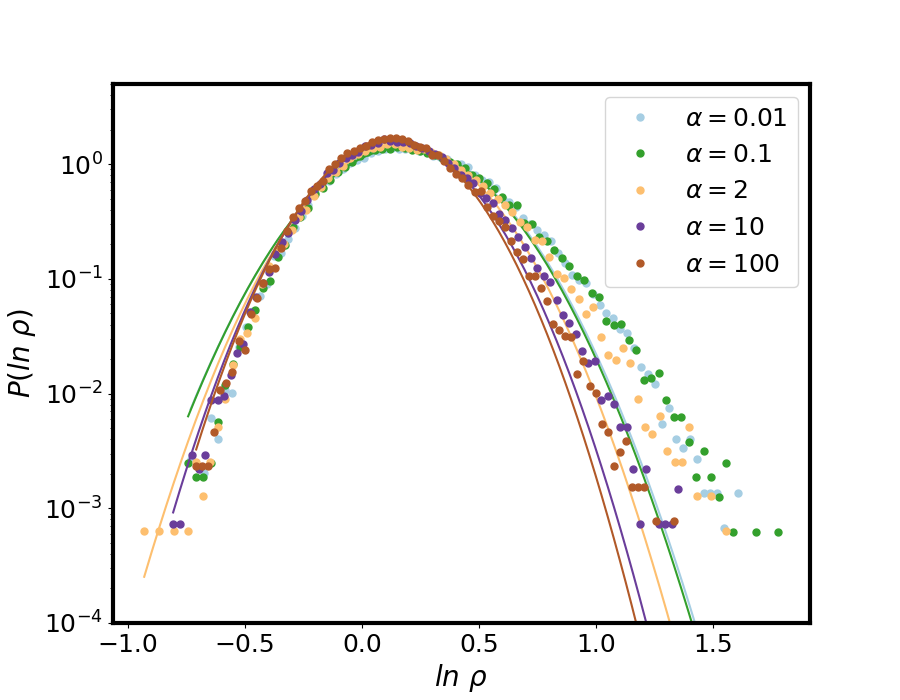}
    \includegraphics[trim={0cm 0cm 1cm 1cm},clip ,width=0.33\linewidth]{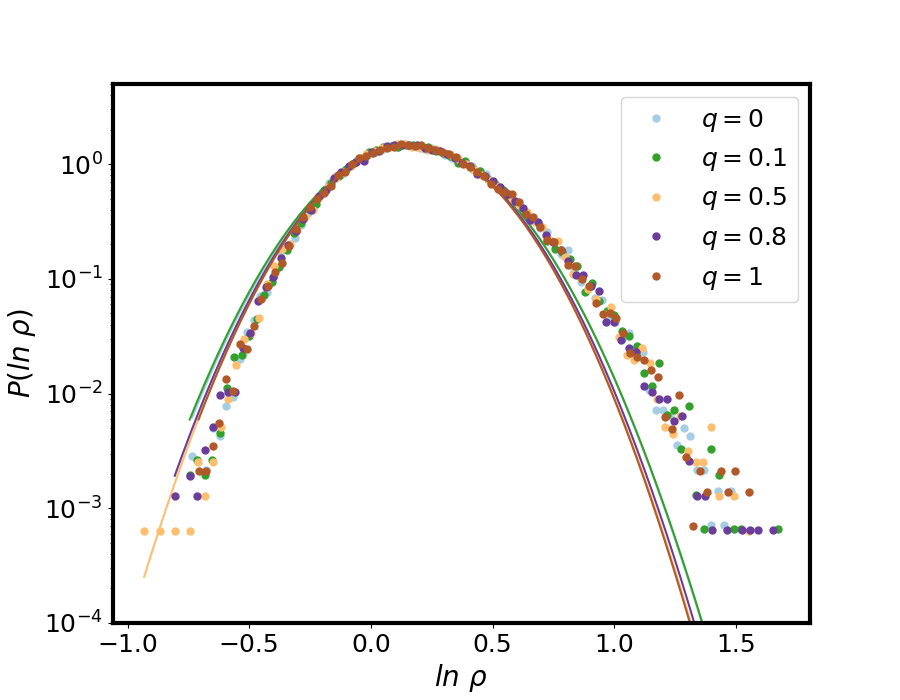}
\caption{PDFs of the magnetic field strength and the density, top and bottom panels, respectively. Columns from left to right show the effect of varying the shock frequency ($\rm p_{shock}$), the steepness of the shock distribution ($\rm \alpha$) and the ratio of fast to slow shocks (q).}
\label{fig:pdfs_parameters}
\end{figure*}

\begin{table*}
\centering
\caption{Diagnostic fits to the magnetic-field PDFs. Here $b=\ln(B/B_0)$.
The Gaussian core is described by $\mu$ and $\sigma$.
The low-field tail is fit as $\ln p(b)=c+\lambda_L b$.
Quoted errors are local histogram/fit errors.
}
\label{tab:model_pdf_fits}
\begin{tabular}{lccccccc}
\hline
Run & $\mu$ & $\sigma$ & $S$ & $\rm K_{\rm exc}$ & $\rm \lambda_L$ & $\rm F_L (\times 10^{-2})$ & $\rm Q_{\rm asym}$ \\
\hline
$\rm \alpha=2,\,p_{shock}=0.01\, q=0$ & & & & & & & \\ \hline
${\mathcal{M}}$=0.5 & $0.001 \pm 0.0001$ & $0.061 \pm 0.001$ & $-0.016$ & $0.017$ & -- & -- & $1.01$ \\
${\mathcal{M}}$=1 & $-0.001 \pm 0.001$ & $0.123 \pm 0.001$ & $0.001$ & $-0.011$ & -- & -- & $1.00$ \\
$\mathcal{M}=2$ & $0.001 \pm 0.001$ & $0.248 \pm 0.002$ & $-0.023$ & $0.010$ & -- & -- & $1.01$ \\
${\mathcal{M}}$=4 & $0.0 \pm 0.004$ & $0.488 \pm 0.005$ & $-0.025$ & $-0.007$ & -- & -- & $1.01$ \\
${\mathcal{M}}$=10 & $-0.001 \pm 0.009$ & $1.224 \pm 0.012$ & $-0.008$ & $0.042$ & -- & -- &  $1.00$ \\
\hline
$\rm \mathcal{M}=2,\, p_{shock}=10^{-2},\,q=0$ & & & & & & & \\ \hline
$\rm \alpha=10^{-2}$ & $-0.281 \pm 0.165$ & $0.505 \pm 0.212$ & $-1.076$ & $1.382$ & $1.744 \pm 0.031$ & $17.71 \pm 0.17$ & $2.27$ \\
$\rm \alpha=10^{-1}$ & $-0.286 \pm 0.126$ & $0.489 \pm 0.151$ & $-1.125$ & $1.608$ & $1.743 \pm 0.023$ & $19.14 \pm 0.18$ & $2.28$ \\
$\rm \alpha=2$ & $-0.252 \pm 0.115$ & $0.409 \pm 0.110$ & $-1.057$ & $1.397$ & $1.871 \pm 0.028$ & $21.84 \pm 0.18$ & $2.13$ \\
$\rm \alpha=10$ & $-0.256 \pm 0.054$ & $0.414 \pm 0.065$ & $-0.945$ & $1.213$ & $2.319 \pm 0.050$ & $15.18 \pm 0.16$ & $1.88$ \\
$\rm \alpha=100$ & $-0.247 \pm 0.044$ & $0.406 \pm 0.056$ & $-0.865$ & $1.036$ & $2.591 \pm 0.059$ & $10.91 \pm 0.14$ & $1.74$ \\
\hline
$\rm \mathcal{M}=2,\, \alpha=2,\,q=0$ & & & & & & & \\ 
\hline
$\rm p_{shock}=0.001$ & $-0.009 \pm 0.003$ & $0.251 \pm 0.004$ & $-1.029$ & $2.970$ & $3.066 \pm 0.177$ & $3.80 \pm 0.08$ & $1.36$ \\
$\rm p_{shock}=0.005$ & $-0.089 \pm 0.027$ & $0.299 \pm 0.026$ & $-1.229$ & $2.196$ & $2.282 \pm 0.066$ & $17.06 \pm 0.17$ & $2.16$ \\
$\rm p_{shock}=0.01$ & $-0.252 \pm 0.115$ & $0.409 \pm 0.110$ & $-1.057$ & $1.397$ & $1.871 \pm 0.028$ & $21.84 \pm 0.18$ & $2.13$ \\
$\rm p_{shock}=0.05$ & $-2.335 \pm 0.025$ & $1.297 \pm 0.033$ & $-0.541$ & $0.292$ & $1.366 \pm 0.035$ & $2.67 \pm 0.07$ & $1.36$ \\
$\rm p_{shock}=0.1$ & $-4.915 \pm 0.019$ & $1.817 \pm 0.024$ & $-0.374$ & $0.148$ & $1.102 \pm 0.057$ & $1.46 \pm 0.05$ & $1.22$ \\
\hline
$\rm \mathcal{M}=2,\, \alpha=2,\,p_{shock}=0.01$ & & & & & & & \\ \hline
$\rm q=0$ & $-0.278 \pm 0.087$ & $0.435 \pm 0.130$ & $-1.061$ & $1.432$ & $1.886 \pm 0.035$ & $1.99 \pm 0.018$ & $2.15$ \\
$\rm q=0.1$ & $-0.214 \pm 0.063$ & $0.440 \pm 0.063$ & $-0.675$ & $1.400$ & $2.030 \pm 0.032$ & $14.12 \pm 0.016$ & $1.74$ \\
$\rm q=0.5$ & $0.062 \pm 0.086$ & $0.557 \pm 0.103$ & $0.321$ & $1.151$ & $2.783 \pm 0.045$ & $3.32 \pm 0.08$ & $0.83$ \\
$\rm q=0.8$ & $0.268 \pm 0.124$ & $0.535 \pm 0.118$ & $0.781$ & $1.148$ & $3.822 \pm 0.269$ & $0.43 \pm 0.02$ & $0.54$ \\
$\rm q=1$ & $0.482 \pm 0.246$ & $0.737 \pm 0.536$ & $0.963$ & $1.035$ & -- & -- & $0.53$ \\
\hline
\end{tabular}
\end{table*}

\section{Characterizing the magnetic field PDFs}

We will describe the magnetic field PDFs with a lognormal characterized by a mean $\rm \mu$ and a variance $\sigma$, plus a tail of the form:
\begin{equation}
    \rm
    \ln P(b)=c+\lambda_L b,
    \label{eq:low_value_fit}
\end{equation}
where $\rm b=ln~(B/B_0)$) and the tail is present below a value $\rm b_L$. $\mu$, $\sigma$, $\rm b_L$, $\rm \lambda_L$, are parameters which will be reported as least-square fits to the PDFs. We will also report the higher moments (skewness, S, and excess kurtosis, K\footnote{Excess with respect to a Gaussian: negative values indicate rarer extrema, and positive values excess extrema with respect to a Gaussian. Zero means equivalent to a Gaussian in the tail distribution.}) of the entire PDF, as well as
\begin{equation}
    \rm 
    F_L=P(b<b_L)
    \label{eq:low_tail_fraction},
\end{equation}
which defines the low-field tail volume-weighted fraction, and
\begin{equation}
    \rm 
    Q_{asym}=(b_{50}-b_5)/(b_{95}-b_{50})
    \label{eq:q_asymm}
\end{equation}
where $\rm b_{5},~b_{50},~b_{95}$, the 5, 50 and 95 percent quantiles of the distribution,
as a measure of low- versus high-side quantile asymmetry.

For completeness and for comparison purposes, we have included two summary plots of the density and magnetic field PDFs from numerical simulations in the Appendix (Fig. \ref{fig:beattie_turb}) using publicly available data of driven MHD turbulence simulations from \citet{Beatie2020}, characterizing them with the above parameters.
In the same Appendix we demonstrate that the asymmetric power law tails are largely unaffected by numerical or physical diffusivity (something hinted at in \citet{Schekochihin_2004} and also privately communicated by Amit Seta from their simulations).

\section{Magnetic field and density PDFs from the model}

We integrate Eqs. \ref{eq:continuity}, \ref{eq:Bhat} and \ref{eq:logB} numerically for about $10^5$ random trajectories, each comprised of 200 steps of size dt=0.01, varying the model parameters between runs. We have checked that the PDFs are converged for this number of trajectories and steps. The size of the time step is chosen according to the total integration time, which equals two turbulence crossing times of a box with a normalized size and sound speed of unity. Running the simulations for more crossing times increases the variance of the lognormal because of the white noise-like random process at the core of the model. The size of the step, dt, is connected to the probability of encountering a shock, $\rm p_{shock}$, since together they determine the average number of shocks per trajectory.

We first look at the density and magnetic field PDFs without shocks ($\rm p_{shock}=0$, Fig. \ref{fig:noshocks}).  Here we have varied only the Mach number, which translates into varying $\rm v_{rms}$ in Eq. \ref{eq:vartheta}. 
The fit parameters of the PDF as described in the previous section, for this and the other model sets, are listed in Table \ref{tab:model_pdf_fits}.
We see that increasing the Mach number leads to a larger variance in both distributions. However, the magnetic field distribution is always slightly wider than that of the density, indicating that the trace of the velocity divergence controls the dynamics instead of the change of magnetic field direction. 
No model parameter explicitly connects to the Alfv\'{e}nic Mach number, so we cannot directly compare to the trend as shown in the right panel of Fig. \ref{fig:beattie_turb}, but, as explained in the previous Section, we expect q to carry some of this dependence, although the functional form is unknown.

The results of varying the remaining parameters ($\rm p_{shock}$,$\alpha$,q), keeping the Mach number equal to 2, are reported in Fig. \ref{fig:pdfs_parameters}. The top panel of the figure shows the effect of varying $\rm p_{shock}$ in the presence of exclusively slow shocks (q=0) and for a fixed Mach number, since we have established that varying $\mathcal{M}$ only increases the variance. 
We notice that, as $\rm p_{shock}$ increases, a low-value tail similar to those observed in simulations becomes increasingly important. The opposite effect (high-value tails) is visible in the density PDF. The fit parameters in Table \ref{tab:model_pdf_fits} confirm the visual impression: varying $\rm p_{shock}$ has two effects: i) the exponential tails become flatter for higher $\rm p_{shock}$ for $\rm p_{shock}$ lower than about $0.1$. ii) as $\rm p_{shock}$ increases further, shocks become frequent enough that they act as an effectively continuous random process more similar to the smooth velocity gradient. As a result, for large $\rm p_{shock}$, the entire distribution shifts to smaller values of $\rm |\mathbf{B}|$ and its variance increases. Since this effect is already present for $\rm p_{shock}=0.1$, we do not display the PDFs for higher values.

Varying the slope of the shock strength distribution ($\alpha$), still with q=0, $\mathcal{M}=2$, and for $\rm p_{shock}=0.01$, we see in the middle panel of the figure that the tails become flatter, and the asymmetry stronger for smaller $\alpha$. More quantitatively, in Table \ref{tab:model_pdf_fits} we see that the slope of the exponential tail steepens, the skewness becomes less negative and the excess kurtosis decreases as $\rm\alpha$ increases.

We also notice in the right panel of the figure that low-value tails 
increasingly dominate as the fraction of compressive (fast) shocks decreases, with the maximum effect for q=0. The opposite effect, e.g. high-value tails, appears for a prevalence of compressive shocks. 

Comparing the values of $\rm \lambda_L$ to those fitted to the PDFs of the FLASH and RAMSES numerical simulations (Appendix \ref{sec:app_a}, Tables \ref{tab:logB_pdf_fits_flash} and \ref{tab:logB_pdf_fits_ramses}), we see that, in the range of parameters we have tried, we reproduce the slopes $\rm \lambda_L\simeq2-3$ that correspond to the highest Alfv\'{e}n Mach numbers in the set, for low shock probabilities ($\rm p_{shock}\lesssim0.005$) or, at fixed shock probability, for the flatter shock strength distributions,  $\alpha\gtrsim2$. Otherwise, the $\rm ln|\mathbf{B}|$ PDFs in simulations with $\mathcal{M_A}\lesssim1$ show  with steeper power-law slopes.
On average, the percentage of the PDF in the tail, $\rm F_L$, is of the same order of magnitude, $\rm F_L\simeq10^{-2}$, as our model, for the highest shock probabilities in the set, $\rm p_{shock}\gtrsim0.05$, otherwise it is, on average, higher in our model than in the MHD simulations.
In terms of the percentage of slow versus fast shocks, there is no real preference in terms of the values of both $\rm \lambda_L$ and $\rm F_L$, with only $\rm q=1$ being excluded because of the complete absence of low-value tails. However, this is also owed to the fact that our fit does not include the high-value tails, which appear for $\rm q\gtrsim0.5$.

\section{Conclusions}

We have argued that in a simple dynamical description of turbulence where we follow a fluid element as it moves in a purely stochastic velocity gradient field, the density PDF will always be lognormal. 
In a similar setup, we have defined a stochastic differential equation for $\rm \ln|\mathbf{B}|$ driven by random stretching and background compressions. In this model, we also introduced intermittent shocks whose rate and magnitude are free parameters.

We show analytically and numerically that introducing these rare events results in PDFs with power-law tails in log space. 
These tails originate in the probability distribution of the shocks, which in our model, and in numerical simulations, is exponential, and therefore leads to a Gamma conditioning of the lognormal. 
We also find that the tails are asymmetric (low-value versus high-value tails) if the effects of shocks are prevalently slow or fast.
Therefore, we propose that compressive MHD turbulence simulations that report this low-value asymmetry are dominated in volume by slow shocks whereas the opposite will be true for simulations that report a high-value asymmetry. 
This behavior can be used as a diagnostic for the types of shocks prevalent in a turbulent flow.

\citet{Rabatin_collins2023} introduced a model for corrections to the lognormal density PDF conceptually similar to the one presented here, in that it looks for deviations from a lognormal in the sampling of a finite number of events. In their model, they treat $\rm ln~\rho$ as the sum of a finite number n of shock-induced increments, and report a PDF with non-Gaussian tails for low n due to finite corrections to the central limit theorem. By contrast, our model is a continuous jump-diffusion description of 
$\rm \ln |\textbf{B}|$ and $\rm ln~\rho$, based the MHD equations where the tails of the PDF are analytically predicted by marginalizing over shock counts. This leads to an explicit connection between the shock sampling and the functional form of the tails.

Considering the above, we propose that the appearance of power-law tails around a core lognormal distribution may be a more general signature of intermittency in turbulence that translates into sampling of rare, Poisson-like events, in an otherwise diffusive mechanism. 

\begin{acknowledgements} EN acknowledges funding from the Italian Ministry for Universities and Research (MUR) through the "Young Researchers" funding call (Project MSCA 000074). EN also acknowledges the Interstellar Institute's program "II7" and the Paris-Saclay University's Institut Pascal for hosting discussions that nourished the development of the ideas behind this work. Particular thanks go to Amit Seta for checking that the low-magnetic field value tails are not caused by Ohmic or numerical diffusivity, to James Beattie for making the turbulent box simulation data available, and to the anonymous referee for recommending additional analyses that sharpened our results.
We are grateful to Andrea Ferrara
for useful insights and to Albert Elias for providing simulation data, which unfortunately
did not end up in this manuscript.
\end{acknowledgements}



\bibliography{markov_processes}
\bibliographystyle{aa}

\begin{appendix}

\onecolumn
\section{Density and magnetic field distributions in numerical simulations}
\label{sec:app_a}

\begin{figure*}
    \centering
\includegraphics[trim={0cm 0cm 0cm 0cm},clip, width=0.45\linewidth]{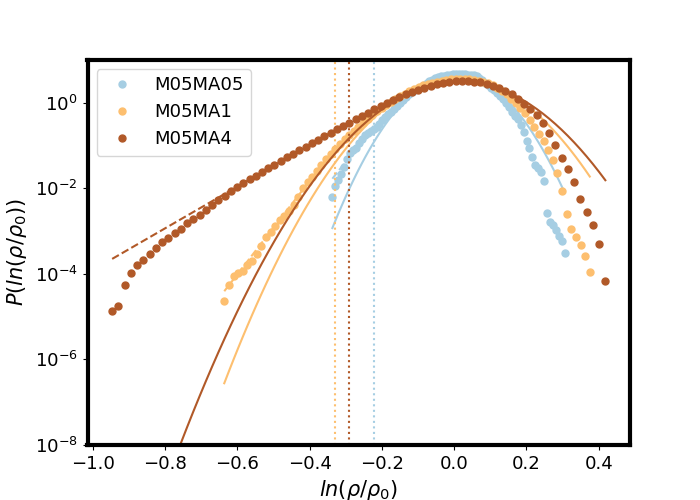}
\includegraphics[trim={0cm 0cm 0cm 0cm},clip, width=0.45\linewidth]{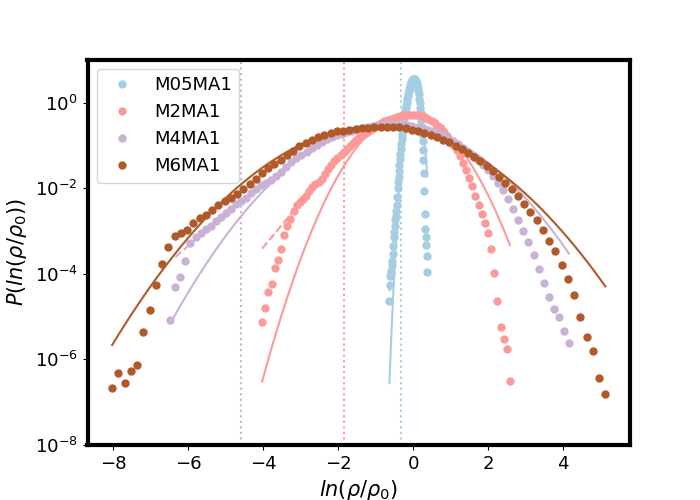}
\includegraphics[trim={0cm 0cm 0cm 0cm},clip, width=0.45\linewidth]{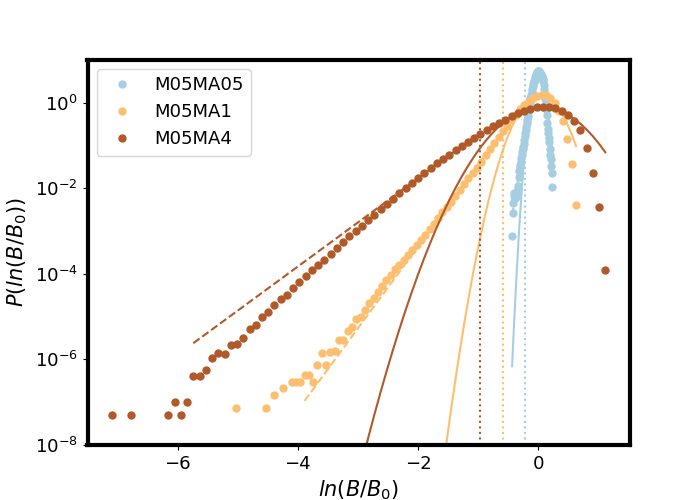}
\includegraphics[trim={0cm 0cm 0cm 0cm},clip, width=0.45\linewidth]{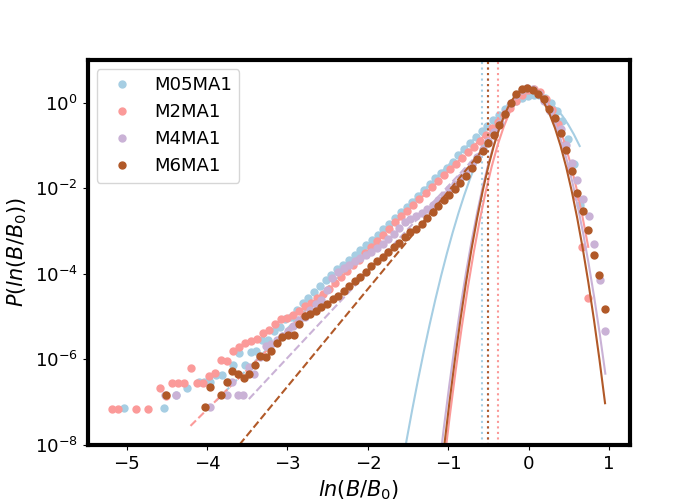}
    \caption{Density(top) and magnetic field strength (bottom) PDFs in driven turbulence simulations with varying Mach number (left) and Alfv\'{e}n Mach number (right) by \citet{Beattie2022}.}
    \label{fig:beattie_turb}
\end{figure*}

\begin{table}
\centering
\caption{Fit parameters for the magnetic field PDFs shown in Fig. \ref{fig:beattie_turb}. 
$\rm \mu$ and $\rm \sigma$ are the fitted lognormal mean and variance. S is the skewness and $\rm K_{\rm exc}$ is the excess kurtosis of the full distribution, $\rm \lambda_L$ is the low-field tail slope from Eq. \ref{eq:low_value_fit}, $F_L$ is the tail volume-weighted fraction from Eq. \ref{eq:low_tail_fraction}, and $b_L$ is the upper edge of the fitted low-field tail. $\rm Q_{asym}$ is the asymmetry from Eq.\ref{eq:q_asymm}. Quoted errors, where shown, are local histogram or fit errors.
The large value of $\rm \lambda_L $ for run M05MA05 indicates the negligible exponential tail in this very low-Alfv\'{e}n Mach number case. 
}
\begin{tabular}{lccccccc}
\hline
\label{tab:logB_pdf_fits_flash}
Run & $\mu$ & $\sigma$ & $S$ & $\rm K_{exc}$ & $\rm \lambda_L$ & $\rm F_L~(\times10^{-3})$ & $\rm Q_{\rm asym}$ \\
\hline
M05MA05 & $0.007 \pm 0.003$ & $0.078 \pm 0.004$ & -0.453 & 0.673 & $19.862 \pm 0.564$ & $7.734 \pm 6.34\times10^{-3}$ & $1.24$ \\
M05MA1 & $0.036 \pm 0.037$ & $0.255 \pm 0.043$ & -1.026 & 2.255 & $4.393 \pm 0.018$ & $48.45 \pm 0.15$ & $1.60$ \\
M05MA4 & $0.052 \pm 0.119$ & $0.482 \pm 0.142$ & -0.997 & 1.534 & $2.371 \pm 0.024$ & $84.32 \pm 0.2$ & $1.74$ \\

M2MA1 & $0.034 \pm 0.011$ & $0.171 \pm 0.011$ & $-1.425$ & $4.380$ & $4.277 \pm 0.021$ & $81.16 \pm 0.2$ & $1.75$ \\
M4MA1 & $-0.008 \pm 0.009$ & $0.173 \pm 0.007$ & $-0.940$ & $4.883$ & $4.625 \pm 0.114$ & $23 \pm 0.1$ & $1.12$ \\
M6MA1 & $-0.019 \pm 0.019$ & $0.166 \pm 0.028$ & $-0.761$ & $3.597$ & $5.247 \pm 0.099$ & $19 \pm 9.9\times10^{-3}$ & $1.07$ \\

\hline
\end{tabular}
\end{table}

\begin{table}
\centering
\caption{Fit parameters for the magnetic field PDFs shown in Fig. \ref{fig:ramses_sims_diffussivity}. All runs have the same Mach number, $\rm\mathcal{M}=2$ and the same Alfv\'{e}nic Mach number $\rm\mathcal{M_A}=2$.}
\label{tab:logB_pdf_fits_ramses}
\begin{tabular}{lccccccc}
\hline
Run & $\mu$ & $\sigma$ & $S$ & $K_{\rm exc}$ & $\lambda_L$ & $F_L (\times10^{-4})$ & $Q_{\rm asym}$ \\
\hline
${\cal{M}}=2$,$256^3$ & $0.010 \pm 0.094$ & $0.676 \pm 0.141$ & -0.833 & 0.851 & $2.816 \pm 0.022$ & $8.755 \pm 0.072$ & $1.68$ \\
${\cal{M}}=2$,$512^3$ & $0.050 \pm 0.065$ & $0.491 \pm 0.081$ & -1.004  & 1.258 & $2.967 \pm 0.015$ & $9.147 \pm 0.026$ & $1.96$ \\
${\cal{M}}=2$,$256^3$,$\eta=0.01$ & $-0.126 \pm 0.040$ & $0.738 \pm 0.072$ & -0.423 & 0.117 & $2.882 \pm 0.040$ & $9.941 \pm 0.076$ & $1.33$ \\
\hline
\end{tabular}
\end{table}

\begin{figure}
    \centering
    \includegraphics[width=0.45\linewidth]{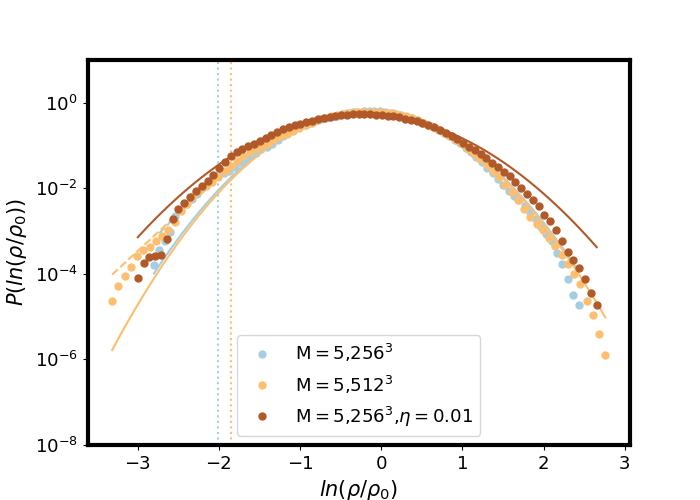}
     \includegraphics[width=0.45\linewidth]{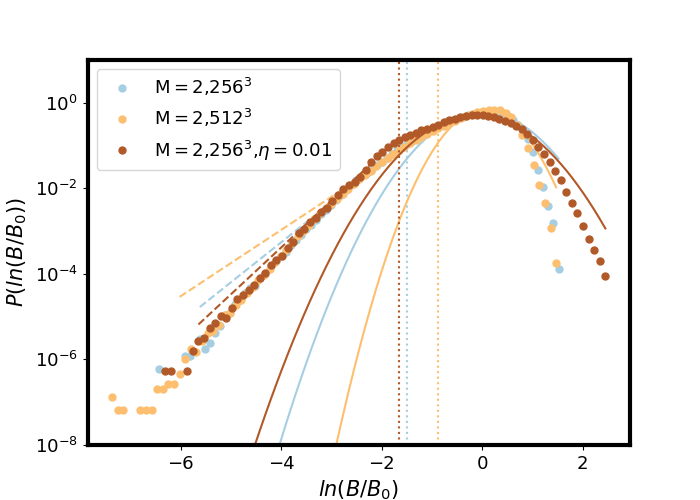}
    \caption{PDFs of $\rm ln~(\rho/\rho_0)$ and $\rm ln~(B/B_0)$ in driven turbulence simulations with the same Mach number ($\rm\mathcal{M}\simeq2$) but different numerical diffussivity (provided by the varying spatial resolution), or explicit Ohmic diffusivity ($\eta=0.01$). The snapshots correspond to one turbulence crossing time. The vertical dotted lines in the right panel indicates $\rm b_L$, which is the fitted value below which the distribution turns to an exponential tail. In this case. $\rm b_L$ is the same for the three PDFs.}
    \label{fig:ramses_sims_diffussivity}
\end{figure}

Figure \ref{fig:beattie_turb} shows typical density and magnetic field distributions from publicly available compressible, ideal, driven MHD turbulence simulations \citep{Beattie2022} performed with the FLASH AMR code \citep{fryxell2000}. We chose this suite of simulations because they cover a wide range of sonic and Alfv\'{e}nic Mach numbers (hereafter $\rm \mathcal{M}$ and $\rm \mathcal{M_A}$, referring always to the rms valiues) at a relatively high resolution.
In the left panel we show the magnetic field distribution for four models with the $\rm \mathcal{M_A}$ but different $\rm \mathcal{M}$, and in the right panel those of four models with the same $\rm \mathcal{M}$ and varying $\rm \mathcal{M_A}$, all taken at the same simulation time. All the PDFs are normalized to the mean value in each snapshot.

The density PDFs are reasonably approximated by lognormals, although asymmetries do exist for high enough $\rm \mathcal{M_A}$ \citep[as noted also by numerous works in the literature, e.g.,][among others]{Hopkins2013,Beattie2022}. 
The magnetic field strength PDFs, however, gradually develop increasingly prominent exponential tails in the low-value part of the distribution with increasing $\rm \mathcal{M_A}$. Other papers have reported the presence of these tails in the positive end of the distribution, e.g. \citet{Schekochihin_2004,Seta_2021}.

The lognormal plus exponential tail fit parameters for these magnetic field PDFs are summarized in table \ref{tab:logB_pdf_fits_flash}. The variance of the lognormal fit increases with increasing $\rm \mathcal{M_A}$, but for this set it does not appear to depend on $\rm \mathcal{M}$. 
The slope of the exponential tail, $\rm \lambda_L$ is the steepest ($\rm \lambda_L\simeq20$) for the lowest  $\rm \mathcal{M_A}$ run in the set, $\rm \mathcal{M_A}=0.5$, which is equivalent to no tail. It is present but steep ($\rm \lambda_L\simeq4-5$) for runs with $\rm \mathcal{M_A}=1$, and more pronounced ($\rm \lambda_L\simeq2.3$) for $\rm \mathcal{M_A}=4$.
The skewness of the entire distribution is always negative (as expected, since the distributions have tails in the low-value part), although without a clear dependence on $\rm \mathcal{M}$ or $\rm \mathcal{M_A}$. The excess kurtosis, $\rm K_{exc}$ is always positive, meaning more extrema with respect to a Gaussian, and shows a weak dependence on $\rm \mathcal{M_A}$. Finally, the values of $\rm Q_{asym}$ show a negative excess for all but the highest $\rm \mathcal{M}$ runs, for which the exponential tails are steeper.

The exponential tails appear to be unaffected by diffussivity. As an example, Fig. \ref{fig:ramses_sims_diffussivity} shows magnetic field PDFs for three additional driven turbulence simulations, aimed at exploring the effect of numerical or Ohmic diffusivity.
We performed these simulations using a custom version of the RAMSES code \citep{teyssier2002,fromang2006} that includes non-ideal MHD effects, as described in \citet{masson12}. Since we only care about the effects of diffusivity, we kept the other parameters constant. We note that the result related to the role of diffussivity persists also for different flow parameters. Here, we assumed $\rm {\mathcal{M}}=2$, $\rm {\mathcal{M_A}}=2$, with an initially uniform magnetic field of $\rm B_0=3~\mu G$ along the x axis. The initial density of the gas was $\rm 10^4~cm^{-2}$ everywhere and it evolved with an isothermal equation of state and a temperature of $\rm T=100~K$.
The rms Mach number of the flow was kept constant by driving turbulence through an Ornstein–Uhlenbeck process, projected into a solenoidal and a compressive component in Fourier space, in a way very similar to the \citet{Beattie2022} simulations illustrated above, and many previous works in the literature. Here we used purely solenoidal driving.

To study the effect of numerical diffusivity, we ran two of these simulations at different resolutions, $256^3$ and $512^3$, keeping everything else the same. The third simulation, at a resolution of $256^3$, includes an explicit Ohmic diffusivity. The Ohmic diffusivity was not calculated self-consistently here because this would entail modeling ionization and other effects relevant to specific astrophysical environments, resulting in small values and physical scales not resolved in this simulation. Instead, we used an artificially large, constant value of $\rm \eta=0.01$ in code units with the sole purpose to see its effects on the PDF. This value sets the effective Reynolds number of this simulation to approximately $\rm \mathcal{R}=200$.

Table \ref{tab:logB_pdf_fits_ramses} summarizes the fit parameters for this second set of simulations. We notice that the two simulations at different resolutions differ in the parameters of the lognormal: The higher resolution simulation has a slightly smaller variance, more negative skewness, and higher excess kurtosis. However, the slope of the low-value tail, the volume fraction of the tail, and the asymmetry are very similar between the two runs, indicating that the tail is largely insensitive to numerical diffusivity.
This is also true of the simulation with Ohmic dissipation: the variance of the lognormal fit is closer to that of the ideal $256^3$ run rather than to that of the $512^3$ run, and its overall skewness and kurtosis are different than both ideal runs. Still, the slope of the low-value tail, the volume fraction of the tail, and the asymmetry are close to the ideal runs. 
Interestingly, the slope of the exponential tail is the same between the three runs consistent with the trend observed in the previous set of simulations: $\rm \lambda_L\simeq 3$ for $\rm \mathcal{M_A}=2$. This is particularly interesting given that the  two sets of simulations were performed using different codes. 
Overall, this behavior supports an ideal MHD origin for the formation of the exponential tail in the magnetic field PDFs.

\section{Fluid parcel trajectories}

In the main text of the paper we presented the PDFs of $\rm ln~\rho$ and $\rm ln~|\mathbf{B}|$ without considering the evolution of individual fluid parcels. Here we show some example time series for some of the models to better illustrate the system dynamics.

\begin{figure}
    \centering
    \includegraphics[width=0.45\linewidth]{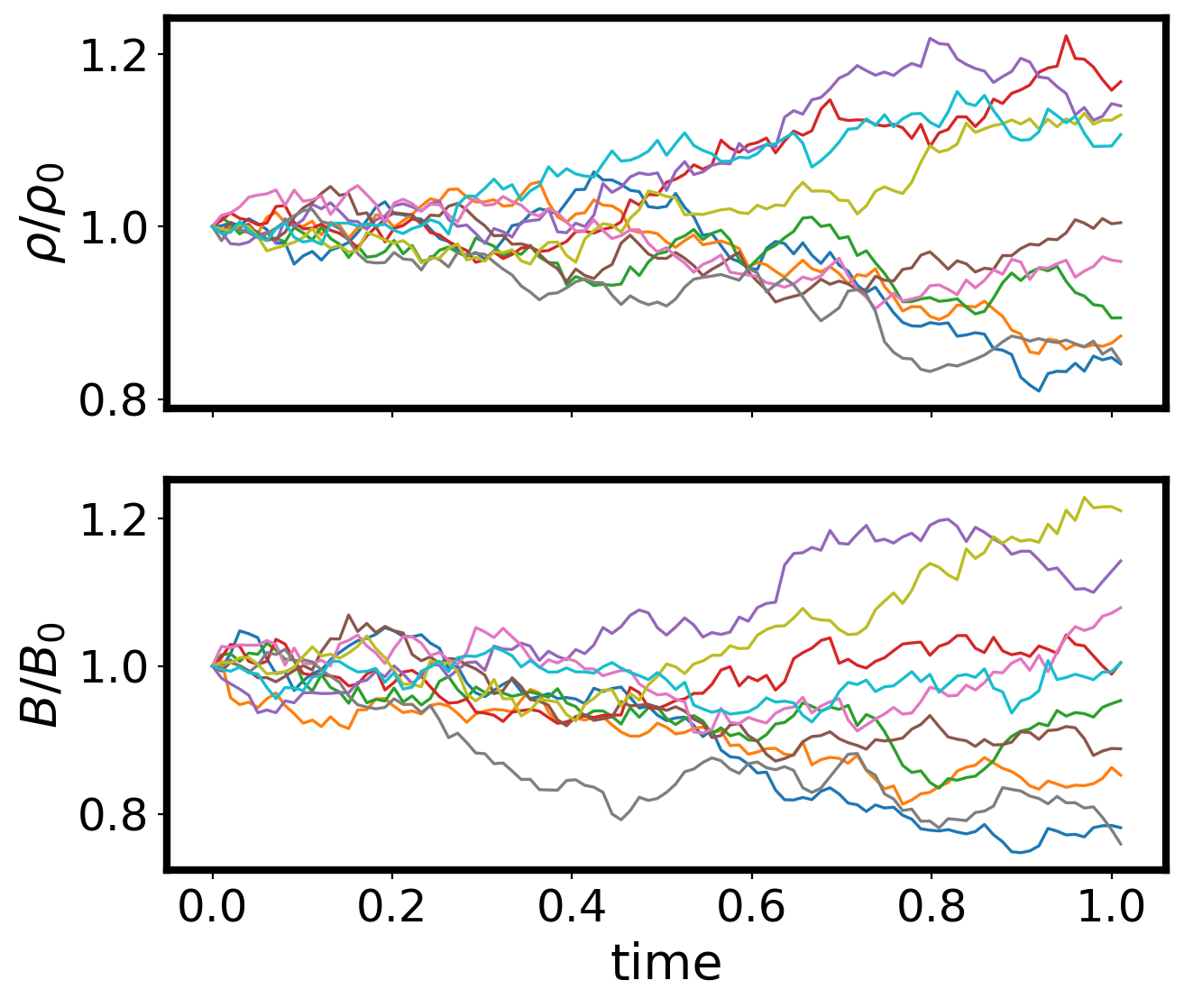}
  \includegraphics[width=0.45\linewidth]{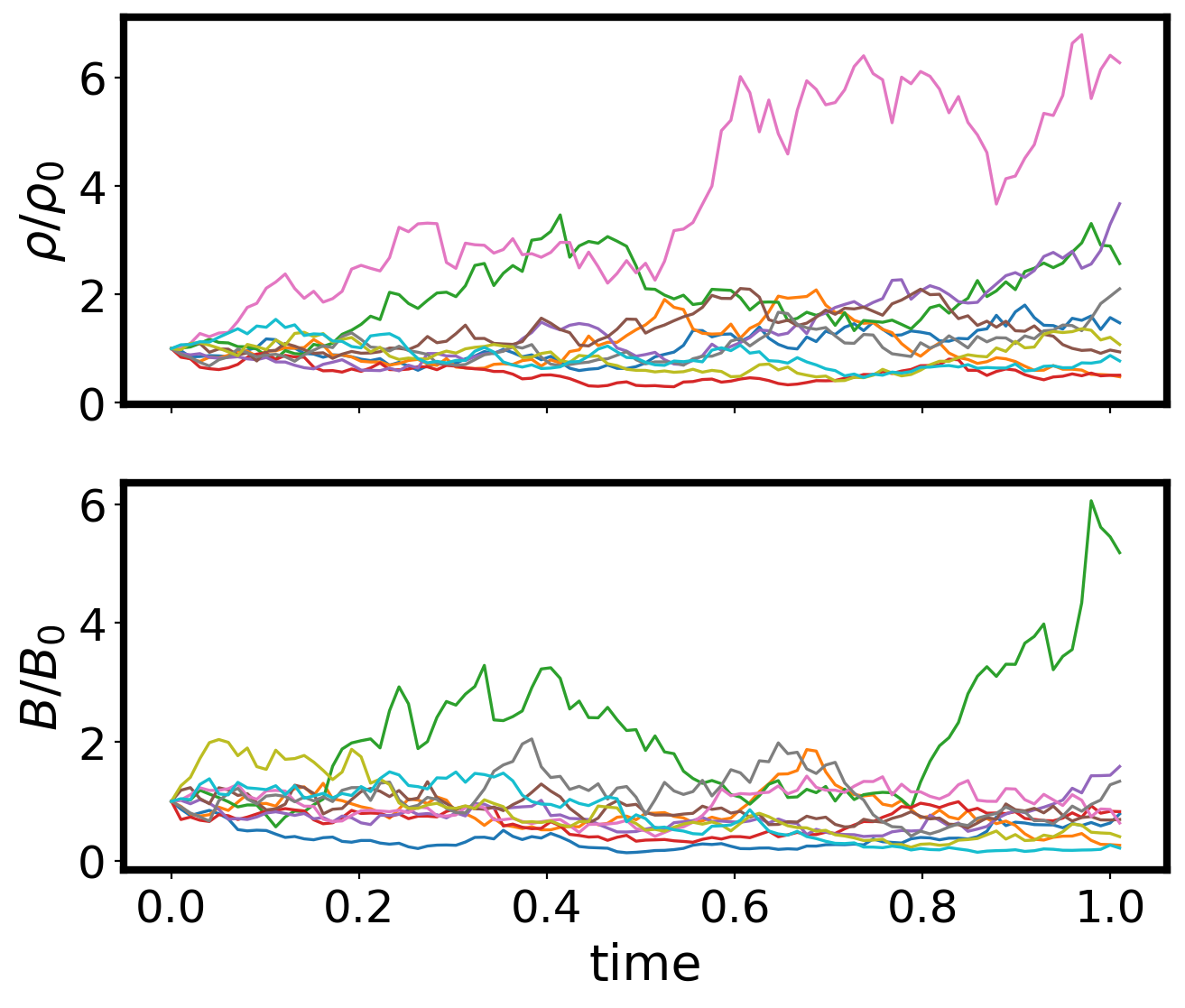}
    \caption{Randomly selected time series of $\rho$ and $\rm |\mathbf{B}|$ in two models with different Mach number ($\mathcal{M}=1$, left, $\mathcal{M}=10$, right), with $\rm p_{shock}=0$, so under pure diffusion.}
    \label{fig:trajectories_machonly}
\end{figure}

Figure \ref{fig:trajectories_machonly} shows time series of $\rho$ and $\rm |\mathbf{B}|$ under pure diffusion ($\rm p_{shock}=0$) for two models with different Mach numbers. The increase in variance with higher $\mathcal{M}$ is clear in the higher increments in both variables.

\begin{figure}
    \centering
    \includegraphics[width=0.45\linewidth]{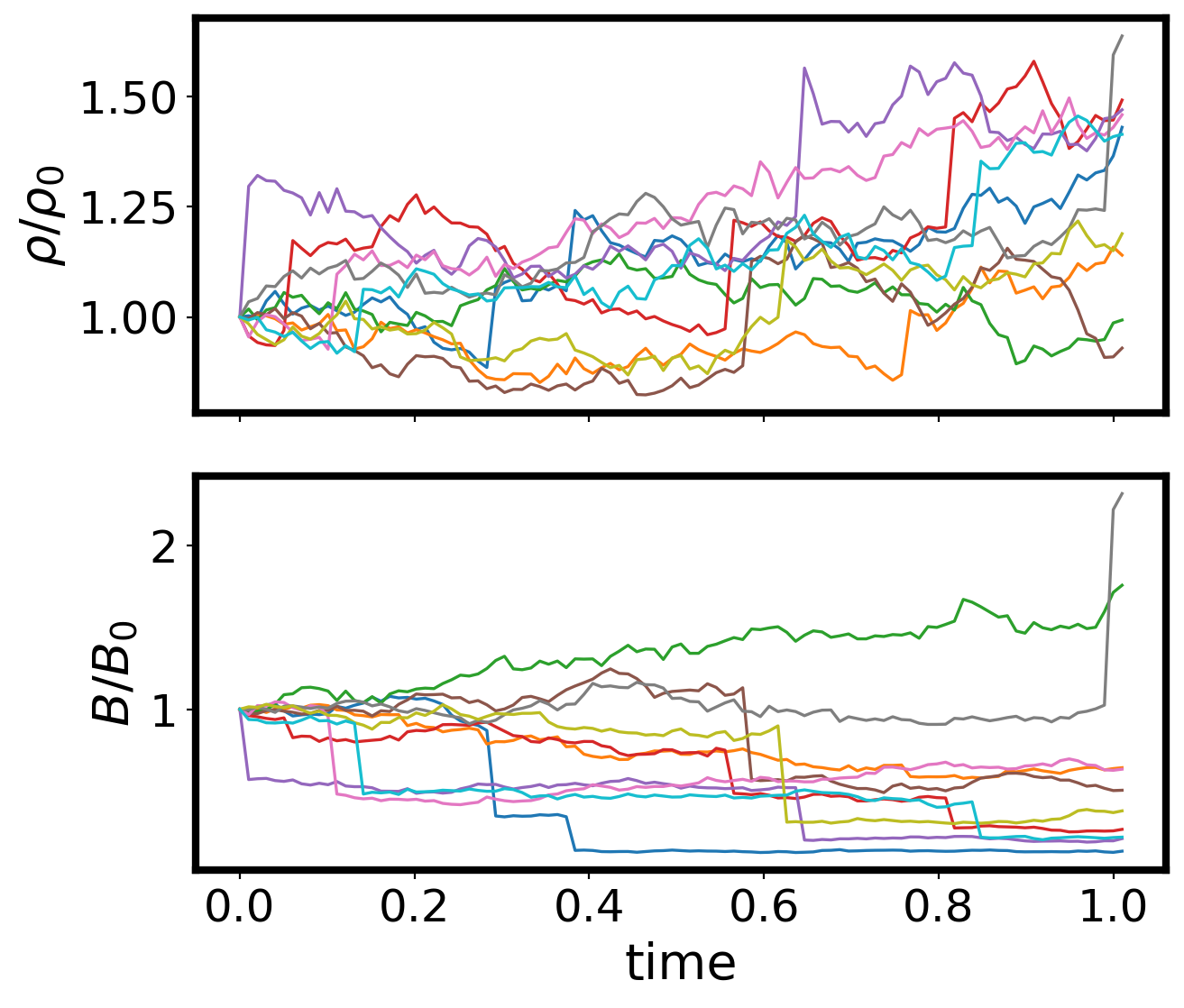}
  \includegraphics[width=0.45\linewidth]{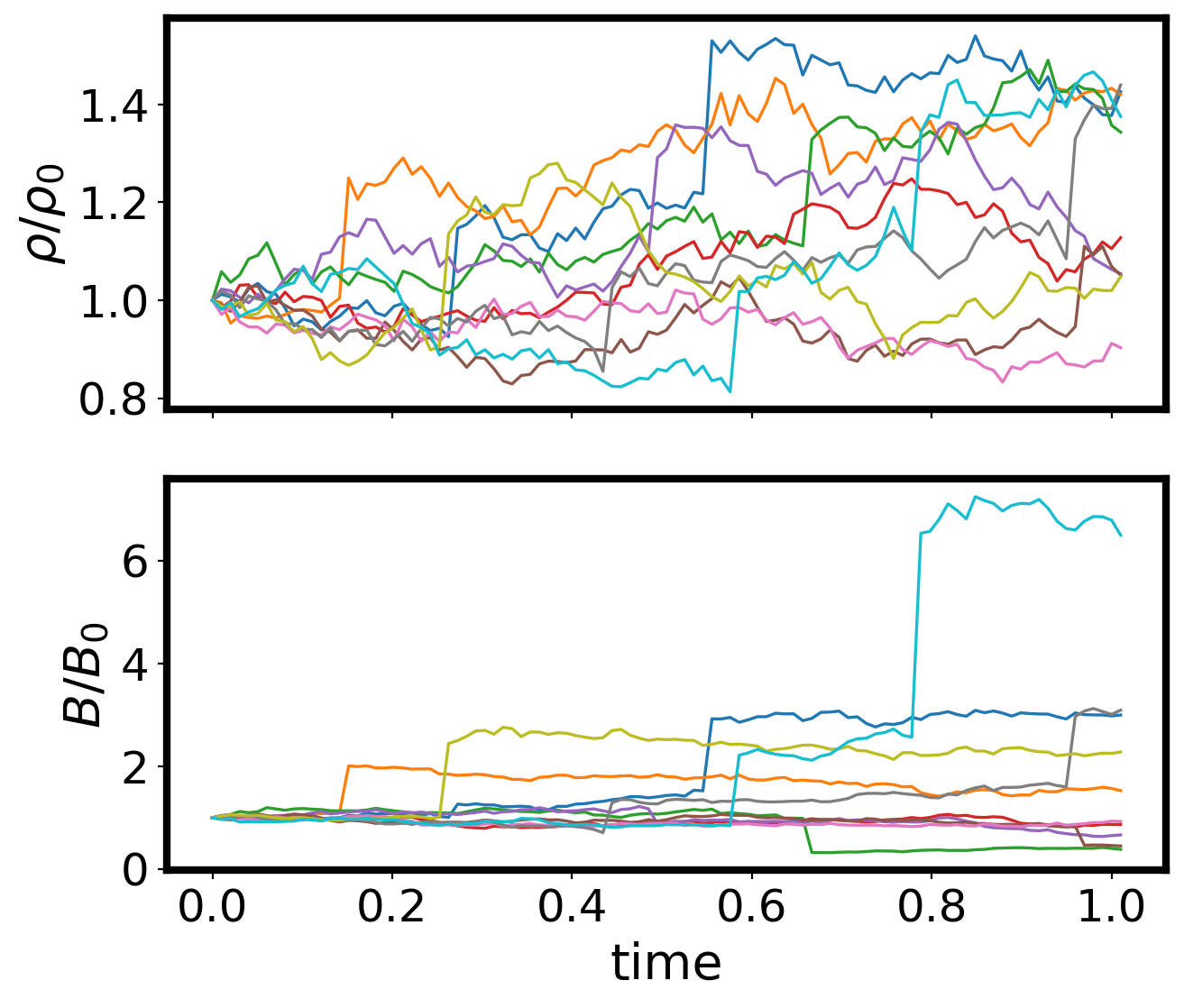}
    \caption{Randomly selected time series of $\rho$ and $\rm |\mathbf{B}|$ in two models with different fast shock percentage ($\rm q=0.1$, left, $\rm q=0.8$, right), with $\rm p_{shock}=0.01$ and $\alpha=2$.}
   \label{fig:trajectories_qdependence}
\end{figure}

The introduction of shocks produces occasional "jumps" in some trajectories, as shown in Fig. \ref{fig:trajectories_qdependence}. The time series clearly show that a higher fraction of fast shocks translates into  more frequent positive jumps in magnetic field strength.

\begin{figure}
    \centering
    \includegraphics[width=0.45\linewidth]{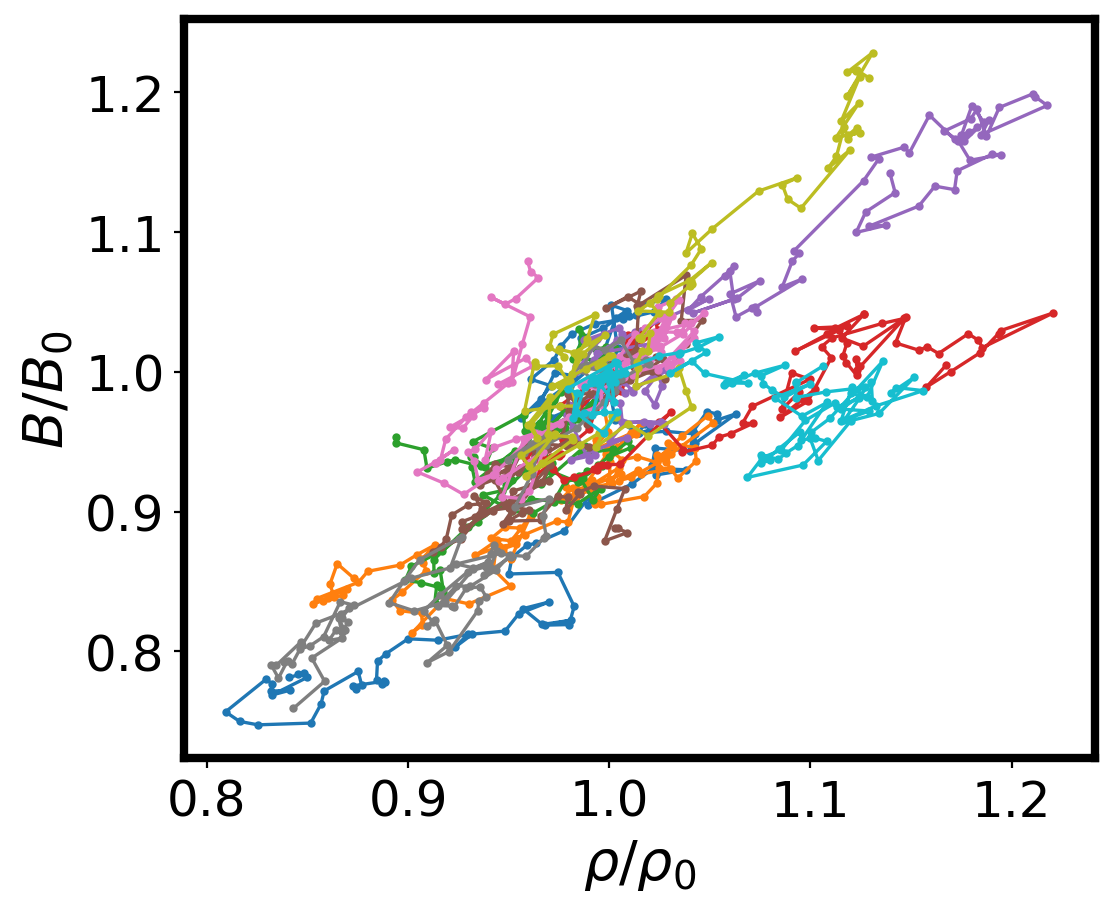}
     \includegraphics[width=0.45\linewidth]{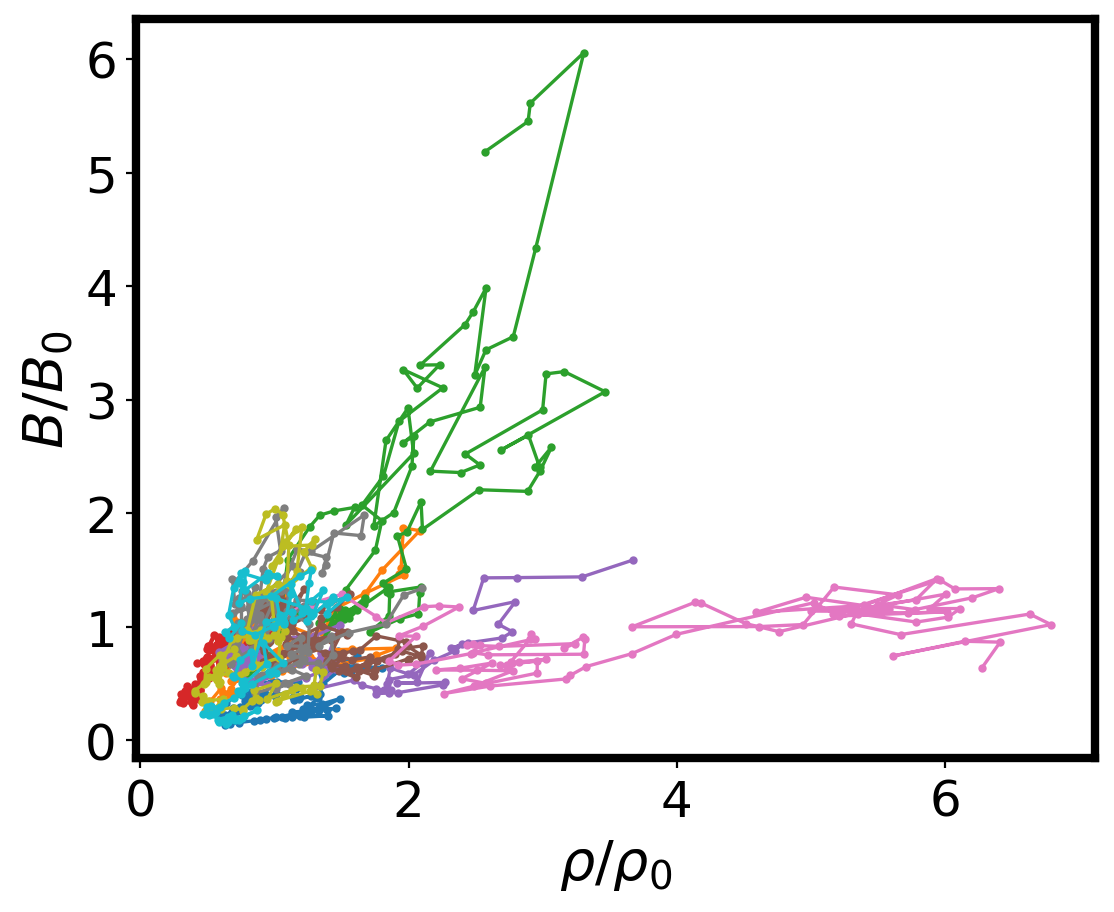}
    \caption{Correlation between $\rm |\mathbf{B}|$ and $\rho$ for the same trajectories shown in Fig. \ref{fig:trajectories_machonly}, with $\mathcal{M}=1$ (left) and $\mathcal{M}=10$ (right).}
    \label{fig:Brho_mach}
\end{figure}

\begin{figure}
    \centering
    \includegraphics[width=0.45\linewidth]{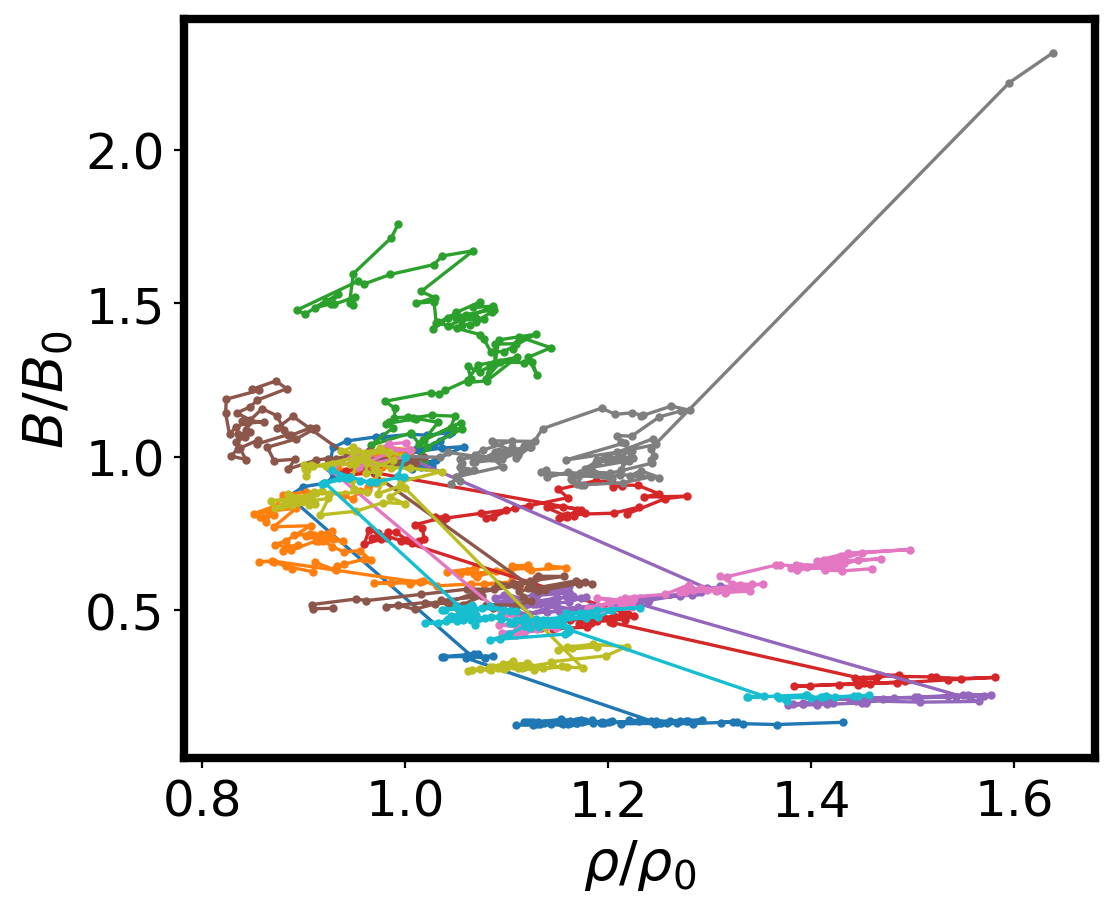}
     \includegraphics[width=0.45\linewidth]{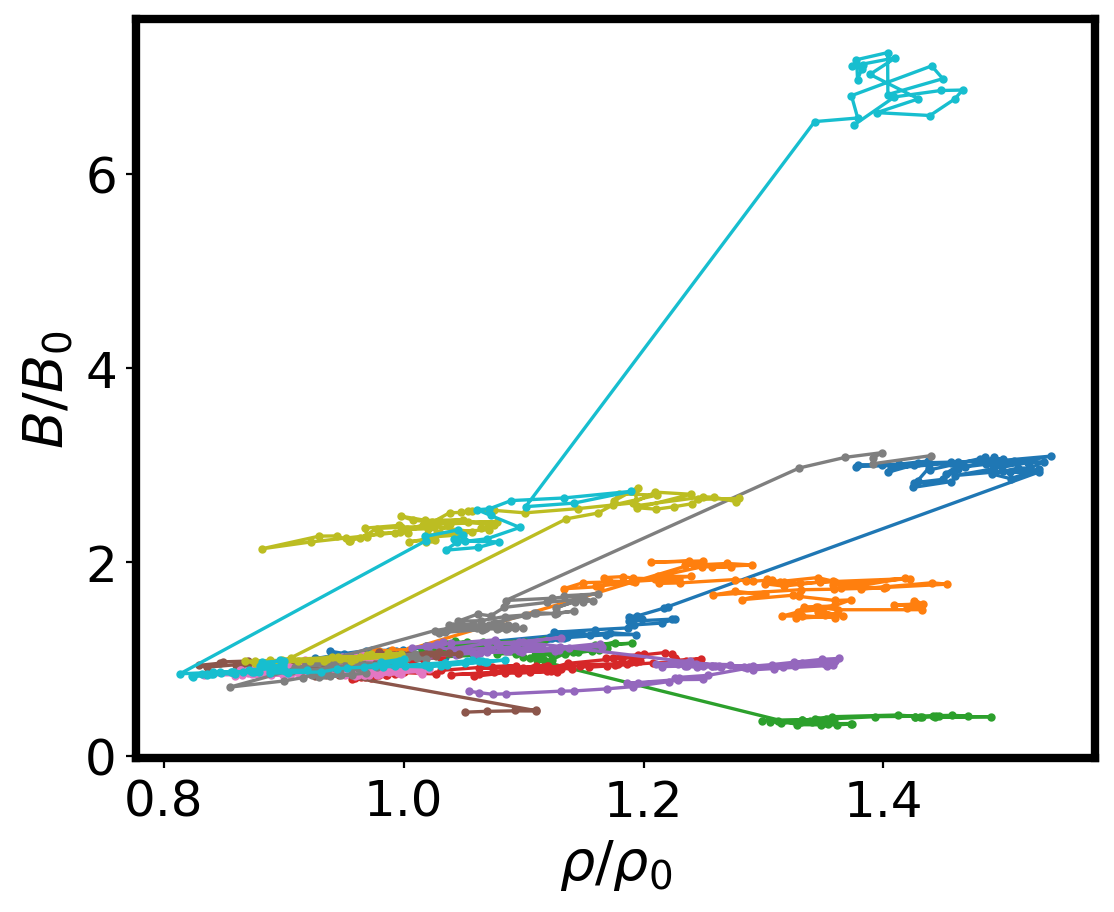}
    \caption{Correlation between $\rm |\textbf{B}|$ and $\rho$ for the same trajectories shown in Fig. \ref{fig:trajectories_qdependence}, with $q=0.1$ (left) and $q=0.8$ (right).}
    \label{fig:Brho_q}
\end{figure}

Finally, we plot the magnetic field versus the density for the same trajectories in Figs. \ref{fig:Brho_mach} and \ref{fig:Brho_q}. We see that the model naturally predicts that the scatter around the $\rm B-\rho$ relation should increase with higher Mach number, and strong outliers away from the mean relation should appear when intermittent shocks are present, as both the mangetic field strenght and the density jump to a new position of the diagram.

\twocolumn

\end{appendix}

\end{document}